\documentclass[12pt]{iopart}
\usepackage[dvips]{graphicx}
\begin{document}
%

\title{Elliptic Flow in Au+Au Collisions at RHIC}

\author {Carla M Vale$^{1}$\footnote{\emph{Present address:} Iowa State 
University, Ames IA 50011} for the PHOBOS Collaboration}
\noindent
B~B~Back$^2$,
M~D~Baker$^3$,
M~Ballintijn$^1$,
D~S~Barton$^3$,
R~R~Betts$^4$,
A~A~Bickley$^5$,
R~Bindel$^5$,
A~Budzanowski$^6$,
W~Busza$^1$,
A~Carroll$^3$,
M~P~Decowski$^1$,
E~Garc\'{\i}a$^4$,
N~George$^{2,3}$,
K~Gulbrandsen$^1$,
S~Gushue$^3$,
C~Halliwell$^4$,
J~Hamblen$^7$,
G~A~Heintzelman$^3$,
C~Henderson$^1$,
D~J~Hofman$^4$,
R~S~Hollis$^4$,
R~Ho\l y\'{n}ski$^6$,
B~Holzman$^3$,
A~Iordanova$^4$,
E~Johnson$^7$,
J~L~Kane$^1$,
J~Katzy$^{1,4}$,
N~Khan$^7$,
W~Kucewicz$^4$,
P~Kulinich$^1$,
C~M~Kuo$^8$,
W~T~Lin$^8$,
S~Manly$^7$,
D~McLeod$^4$,
A~C~Mignerey$^5$,
M~Ngyuen$^3$,
R~Nouicer$^4$,
A~Olszewski$^6$,
R~Pak$^3$,
I~C~Park$^7$,
H~Pernegger$^1$,
C~Reed$^1$,
L~P~Remsberg$^3$,
M~Reuter$^4$,
C~Roland$^1$,
G~Roland$^1$,
L~Rosenberg$^1$,
J~Sagerer$^4$,
P~Sarin$^1$,
P~Sawicki$^6$,
W~Skulski$^7$,
P~Steinberg$^3$,
G~S~F~Stephans$^1$,
A~Sukhanov$^3$,
J-L~Tang$^8$,
M~B~Tonjes$^5$,
A~Trzupek$^6$,
G~J~van~Nieuwenhuizen$^1$,
R~Verdier$^1$,
G~Veres$^1$,
F~L~H~Wolfs$^7$,
B~Wosiek$^6$,
K~Wo\'{z}niak$^6$,
A~H~Wuosmaa$^2$ and
B~Wys\l ouch$^1$

\vspace{3mm}

\small
\noindent
$^1$~Massachusetts Institute of Technology, Cambridge, MA 02139-4307, USA\\
$^2$~Argonne National Laboratory, Argonne, IL 60439-4843, USA\\
$^3$~Brookhaven National Laboratory, Upton, NY 11973-5000, USA\\
$^4$~University of Illinois at Chicago, Chicago, IL 60607-7059, USA\\
$^5$~University of Maryland, College Park, MD 20742, USA\\
$^6$~Institute of Nuclear Physics PAN, Krak\'{o}w, Poland\\
$^7$~University of Rochester, Rochester, NY 14627, USA \\
$^8$~National Central University, Chung-Li, Taiwan 

\ead{cmvale@bnl.gov}

\begin{abstract}
Elliptic flow is an interesting probe of the dynamical evolution of the dense 
system formed in the ultrarelativistic heavy ion collisions at the 
Relativistic Heavy Ion Collider (RHIC). The elliptic flow dependences on 
transverse momentum, centrality, and pseudorapidity were measured using data 
collected by the PHOBOS detector, which offers a unique opportunity to study 
the azimuthal anisotropies of charged particles over a wide range of 
pseudorapidity. These measurements are presented, together with an overview of 
the analysis methods and a discussion of the results.
\end{abstract}

Collisions of Au nuclei at the Relativistic Heavy Ion Collider (RHIC) are 
the most energetic nucleus-nucleus collisions ever achieved in a laboratory,
allowing for the study of the properties of nuclear matter under extreme 
conditions. In order to characterize the medium created in the Au+Au 
collisions at RHIC, it is necessary to establish that the particles in the 
system undergo enough reinteractions to reach a state of thermal equilibrium. 
Only in such a state may the evolution of the medium be described in terms of 
thermodynamical quantities. Elliptic flow is one of the experimental 
observables than can help resolve this question. It originates in non-central 
collisions, where due to the incomplete overlap of the nuclei, the region of
overlap is not azimuthally symmetric. If the medium happens to behave 
collectively, this initial asymmetry will give rise to pressure gradients that
change with the azimuthal angle and will modify the angular distribution of 
the produced particles. Elliptic flow, or $v_2$, measures the amplitude of the 
azimuthal anisotropy in the observed particle distributions, and a strong 
signal suggests that such pressure gradients occurred early in the evolution 
of the system.

\begin{figure}[t]
\begin{center}
\includegraphics[width=12cm]{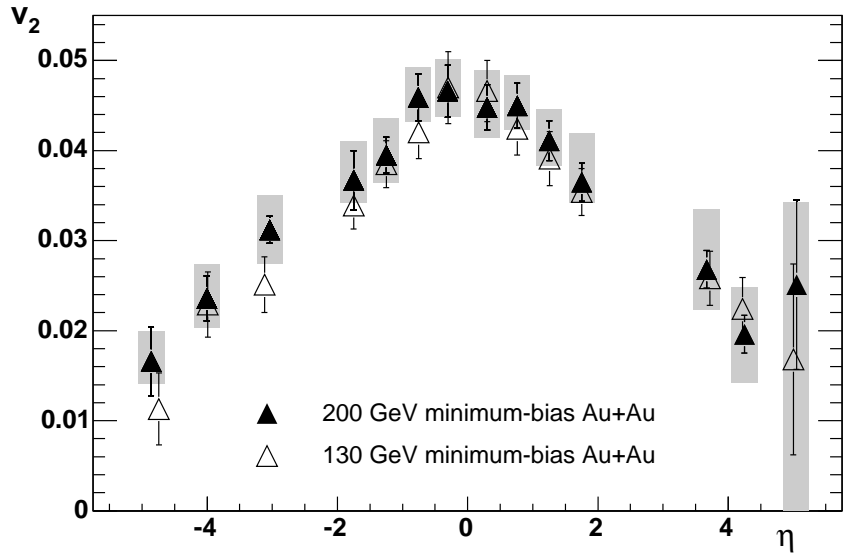}
\end{center}
\caption{\label{fig:v2eta_en} Pseudorapidity dependence of $v_2$ for charged
hadrons in minimum bias Au+Au collisions at $\sqrt{s_{NN}}=$130~GeV (open 
triangles) \cite{phflow130} and 200~GeV (solid triangles) 
\cite{Phobos_flow_200}. The gray boxes around the 200~GeV data points 
represent the systematic errors.}
\end{figure}

Two analysis methods are employed in PHOBOS for the measurement of flow. The 
hit-based analysis method, described in \cite{phflow130}, was first used for 
the analysis of 130~GeV data, providing results on the centrality and 
pseudorapidity dependence of $v_2$. For the analysis of 200~GeV data, the same 
method was applied without changes. The pseudorapidity dependence of $v_2$ for
minimum bias charged hadrons at both energies is shown in figure 
\ref{fig:v2eta_en}. At the startup of RHIC, the expectation was that the 
pseudorapidity dependence of $v_2$, if any existed, should be weak. This
result, showing a strong dependence, was very surprising at the time and still 
poses a challenge to theoretical models. In order to further investigate the 
topic of elliptic flow, a new method which uses tracks in the PHOBOS 
spectrometer, was developed and applied to the 200~GeV data 
\cite{Phobos_flow_200}. That method extends the previous measurements, and 
offers an independent verification of the hit-based analysis results.

The PHOBOS detector is based on silicon pad technology. Several detector 
systems are used for measurements based on hits (single layer detectors), or 
based on tracks (multiple-layer vertex detector and spectrometer). The two 
spectrometer arms are placed inside a dipole magnetic field of 2 T, to allow 
for momentum measurements. The single layer detectors provide a wide 
pseudorapidity coverage, with the octagonal multiplicity detector covering the 
range $|\eta| < 3.2$ (for collisions at the nominal interaction point), and 
two sets of three silicon ring detectors covering a more forward region: 
$3.0 < |\eta| < 5.4$. In these detectors, the passage of particles is recorded 
by their energy deposition in the silicon (a ``hit''). 
A more detailed description of the PHOBOS detector system can be found in 
\cite{phobos_det}.

The experimental trigger is provided by two sets of scintillator detectors, 
the paddle counters, located on each side of the interaction region, along the 
beam line. Additionally, a vertex trigger was given by two sets of \v{C}erenkov
detectors. The vertex trigger was crucial in obtaining the data set used by 
the hit-based analysis, which requires events with collision vertices centered 
at $z = - 34$~cm, where $z$ is the beam axis. 

\begin{figure}[t]
\begin{center}
\includegraphics[width=12cm]{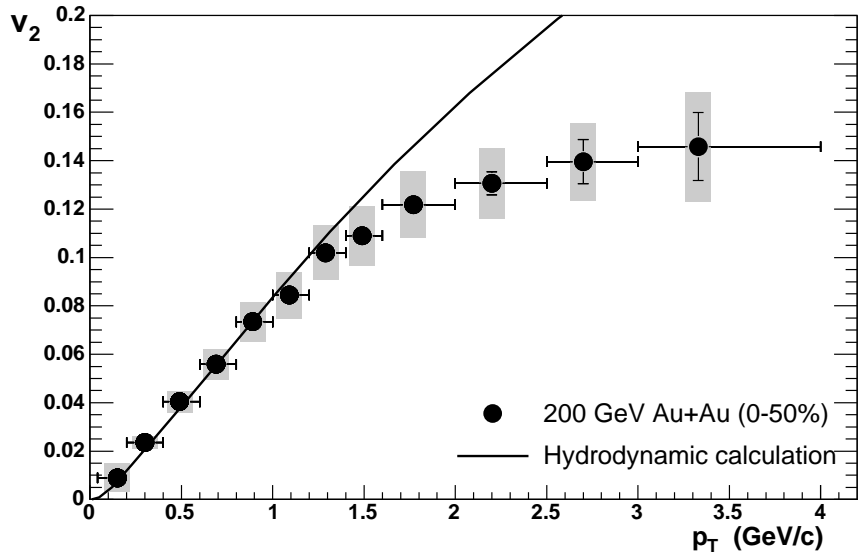}
\end{center}
\caption{\label{fig:v2pt} Elliptic flow of charged hadrons, measured as a 
function of transverse momentum, for the most central 50\% of the 200 GeV 
Au+Au cross-section \cite{Phobos_flow_200}. The gray boxes show the 
systematic uncertainty. The pseudorapidity range of this measurement is 
$0 < \eta < 1.5$. The line corresponds to a hydrodynamical model 
calculation \cite{Huovinen}.}
\end{figure}

The new track-based method determines $v_2$ by correlating the azimuthal 
angles of tracks reconstructed in the spectrometer with the event plane 
measured using the multiplicity detector. Unlike the hit-based analysis, 
the track-based method uses events with vertices near the nominal interaction 
region, in order to maximize the track acceptance in the spectrometer. Within 
this vertex range (-8~cm$<v_z<$10~cm),
the central region of the octagon detector is not azimuthally symmetric, due 
to 4 gaps designed to let particles in the spectrometer and vertex detectors' 
acceptances to go through without traversing additional material. Therefore, in
order to have full azimuthal acceptance for the event plane determination, 
only parts of the detector further away from the nominal interaction region, 
which have no gaps, were used. Following the strategy outlined in 
\cite{Poskanzer}, two subevents of equal multiplicity, and symmetric over 
pseudorapidity, are employed to determine both the event plane and its 
resolution. A detailed description of this analysis method can be found in 
\cite{cmv}.

Figure \ref{fig:v2pt} shows the measured transverse momentum dependence of 
the elliptic flow amplitude in the range $0 < \eta < 1.5$, for events in the 
top 50\% centrality. A curve representing a hydrodynamical model prediction 
\cite{Huovinen}, for the same event class and phase space cuts, is also shown.
The $v_2$ is seen to grow almost linearly as a function of transverse momentum,
up to $p_T \sim 1.5$~GeV/c, and then it appears to saturate at a value of about
0.14. This shape of the $v_2(p_T)$ curve has been observed previously, both for
charged hadrons and identified particles (see, for example: \cite{Star_v2_vsnpart_vspt_130, Phenix_v2_twopartcor_130, Star_v2_Pid_130, Phenix_v2_pid_200}).

The contribution of non-flow effects to experimental measurements of $v_2$ has
been discussed extensively \cite{BDO1, BDO2, Star_v2_Cumulant_130, KandT}. 
In this analysis, the expectation is that by determining the event plane using 
two well separated sub-events, which are also separated from the 
pseudorapidity region where $v_2$ is being measured, the contribution of such 
non-flow effects should be noticeably reduced, particularly 
those originating from short range two-particle correlations. By comparing 
the result of figure \ref{fig:v2pt} with results from reference 
\cite{Star_v2_Cumulant_130}, obtained using reaction plane and 2 and 
4-particle cumulant methods, it is seen that the PHOBOS result shows best 
agreement with the STAR result obtained with the 4-particle cumulant method, 
implying that the PHOBOS result is indeed less sensitive to non-flow effects.

\begin{figure}[ht]
\begin{center}
\includegraphics[width=11cm]{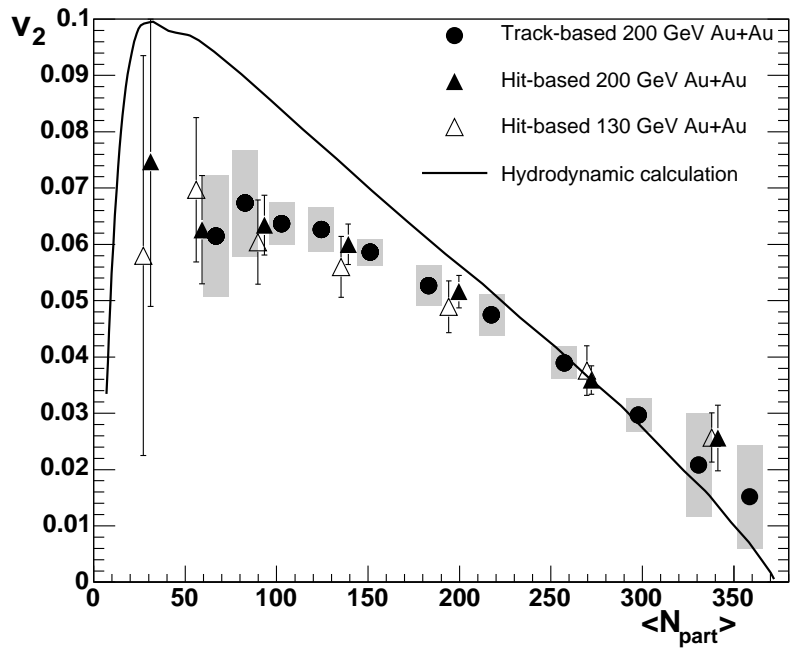}
\end{center}
\caption{\label{fig:v2npart} Centrality dependence of the elliptic flow, for 
$|\eta| < 1$, shown as a function of the number of participants $N_{part}$ 
\cite{Phobos_flow_200}. 
The closed circles represent the result of the track-based method, and the 
closed triangles show the result of the hit-based method, both for collisions 
at 200~GeV. The open triangles show the previously obtained result from the 
hit-based method at 130~GeV \cite{phflow130}. The gray boxes represent the 
systematic 
uncertainties for the track-based method. The line shows a hydrodynamical 
calculation at $\sqrt{s_{NN}} = 200$~GeV \cite{Huovinen}.}
\end{figure}

The results for the centrality dependence of $v_2$, obtained with both 
analysis methods, are compared in figure \ref{fig:v2npart}. Also included in 
the figure is the result for 130~GeV, previously obtained with the hit-based 
method \cite{phflow130}. A very good agreement between the two methods is seen.
The curve shown in the figure corresponds to the same hydrodynamical model 
mentioned previously. 
  
\begin{figure}[ht]
\begin{center}
\includegraphics[width=10cm]{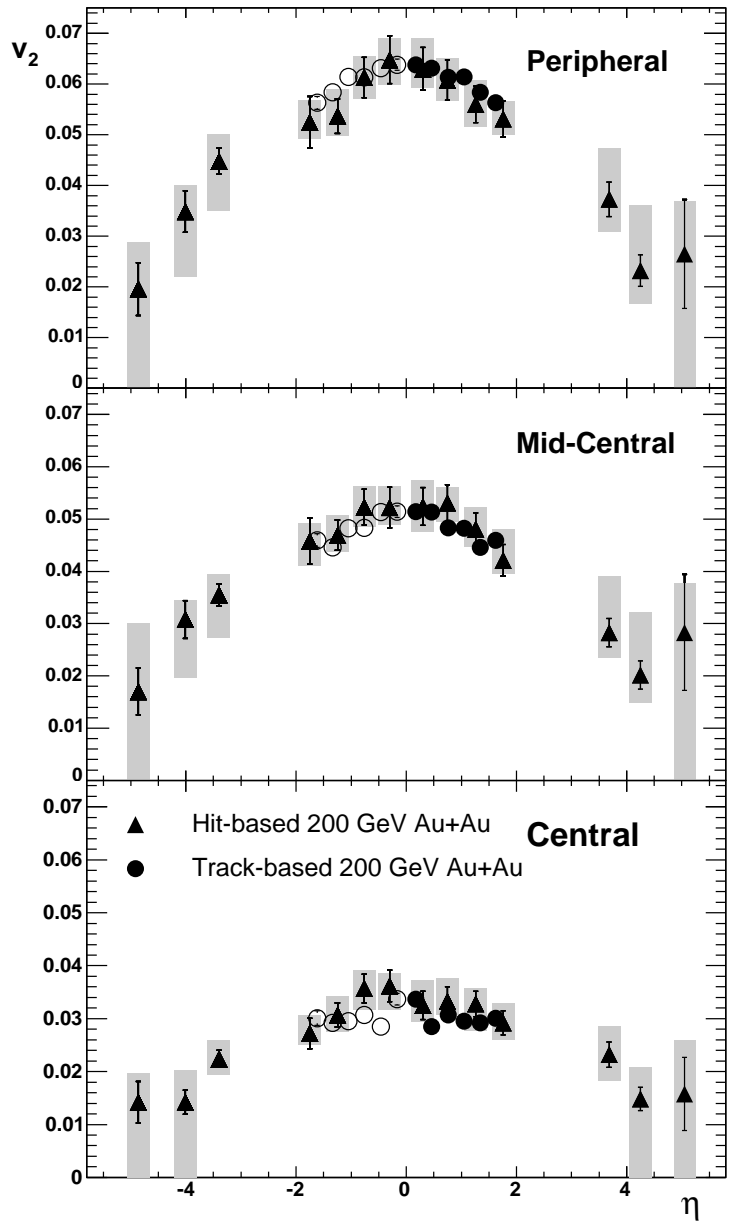}
\end{center}
\caption{\label{fig:v2eta} Elliptic flow of charged hadrons measured as a 
function of pseudorapidity \cite{Phobos_flow_200}, for three centrality 
classes: $25-50\%$, $15-25\%$ and $3-15\%$, respectively from top to 
bottom (top is peripheral, bottom 
is central). The hit-based method results are represented by triangles, and the
track-based results are shown by circles, with open circles representing a 
reflection of the results about mid-rapidity. The gray boxes indicate the 
systematic errors for the hit-based method results.}
\end{figure}

\begin{figure}[ht]
\begin{center}
\includegraphics[width=12cm]{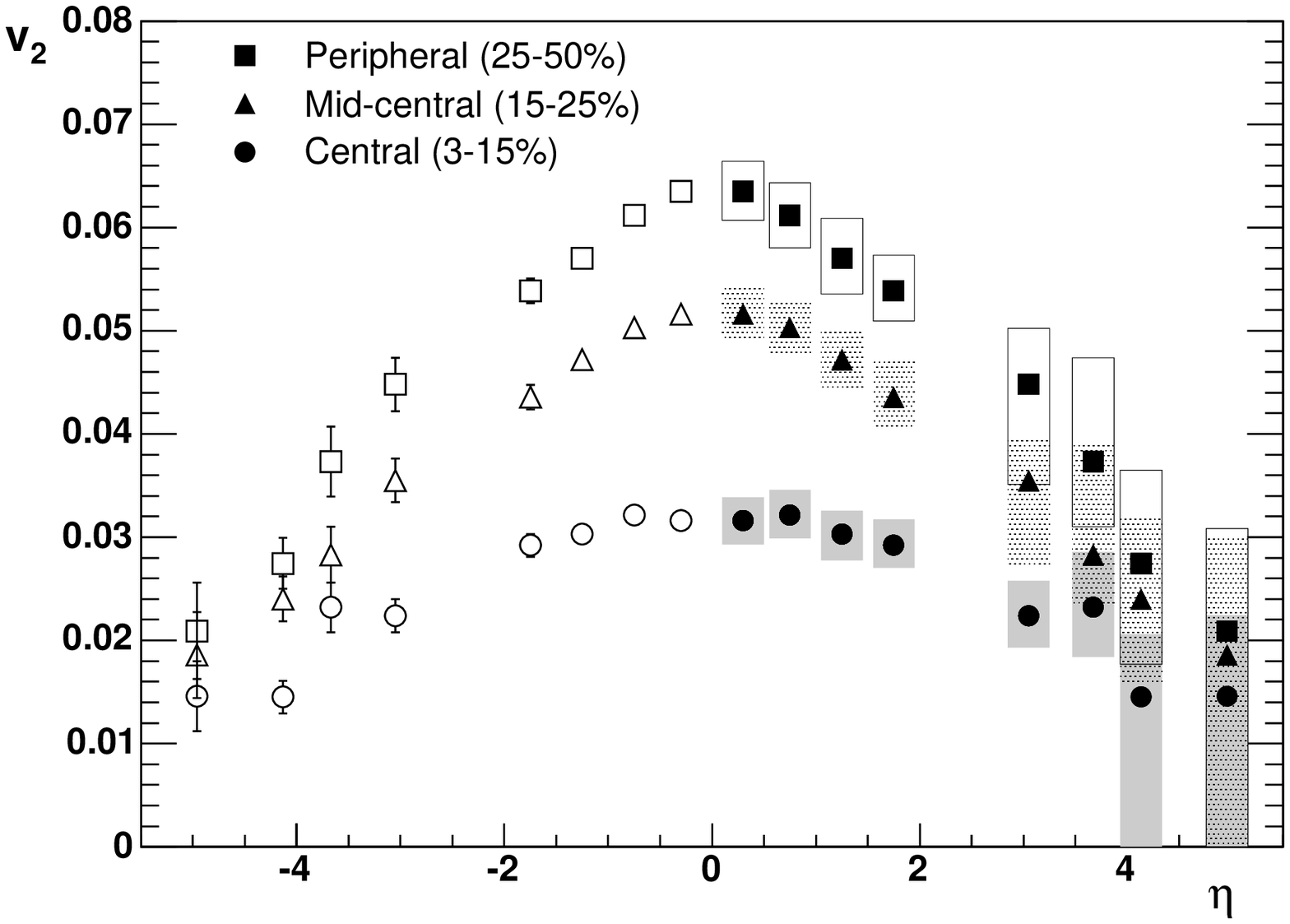}
\end{center}
\caption{\label{fig:v2comb} Pseudorapidity dependence of the charged hadron 
elliptic flow for 3 centrality regions ($25-50\%$ squares, $15-25\%$ triangles,
$3-15\%$ circles), for 200~GeV Au+Au collisions \cite{Phobos_flow_200}. The 
results for positive $\eta$ were obtained by reflecting the hit-based 
measurements about 
mid-rapidity, and then combining them with the track-based measurements. The 
systematic errors are shown as boxes for $\eta > 0$, and the statistical errors
are shown for the reflected points (open symbols) at negative $\eta$.}
\end{figure}

The shape of $v_2(\eta)$ for three centrality classes (``peripheral'': 
$25-50\%$, ``mid-central'': $15-25\%$, ``central'': $3-15\%$) is shown in 
figure \ref{fig:v2eta}, with results from both analysis methods. The agreement 
between the two methods, within their range of overlap, is very good. The 
general features of the shape are common to the three centrality classes, and 
resemble that of the minimum bias result from 
figure~\ref{fig:v2eta_en}. 

Figure~\ref{fig:v2comb} shows $v_{2}(\eta)$ for the same three centrality 
regions, but with the hit-based and track-based methods' results combined. 
Given that these methods employ different detectors and techniques, they are 
taken as independent, and as such, it is assumed when combining the results 
that their errors are not correlated. The combination procedure is detailed in 
\cite{Phobos_flow_200}. At least for the semi-peripheral collisions, it is 
clearly seen that the decrease of $v_2$ with $\eta$ starts already near 
mid-rapidity. The shape of the measured pseudorapidity dependence of $v_2$ 
appears to change only by a scale factor as centrality increases, but a 
flattening of the shape for the most central bin studied cannot at present be 
ruled out.

\begin{figure}
\begin{center}
\includegraphics[width=11cm]{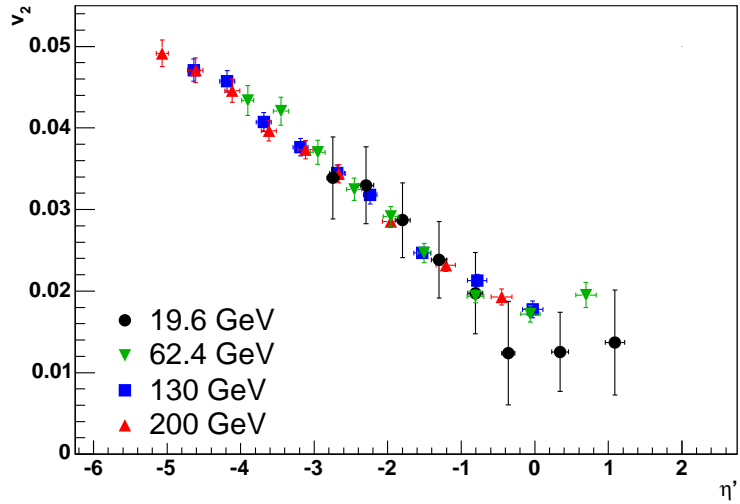}
\end{center}
\caption{\label{fig:limfrag} Elliptic flow in Au+Au collisions for four beam 
energies, averaged over centrality (0-40\%), and show as a function of 
\mbox{$\eta' = |\eta| - y_{beam}$} \cite{phlimfrag}. The error bars represent 
the statistical errors only.}
\end{figure}

Another recent result obtained by PHOBOS \cite{phlimfrag}, with a hit-based 
method similar to the one mentioned above, used data sets at four different 
RHIC beam energies to compare the pseudorapidity dependence of the elliptic 
flow within the context of limiting fragmentation. The results, presented in
figure \ref{fig:limfrag}, are shown as a function of $\eta'$, which is the 
value of the pseudorapidity when shifted by the beam rapidity. It is seen that 
the elliptic flow at all four energies studied appears to be independent of 
energy in $\eta'$, displaying limiting fragmentation throughout the entire 
range of $\eta'$. This is another surprising feature of elliptic flow results
at RHIC, given that particle production in the limiting fragmentation region 
is thought generally to be distinct from that at midrapidity, but in this case
there is no evidence for two separate regions in any of the four energies 
analyzed.

In conclusion, the PHOBOS collaboration has presented an ensemble of 
measurements of charged hadron elliptic flow, including a unique measurement 
of the pseudorapidity dependence of $v_2$ over a large range for several 
beam energies. Both $v_2(p_T)$ and $v_2(N_{part})$ are well described by 
hydrodynamical models, within their range of applicability (i.e., mid-central 
to central collisions, and $p_T$ 
up to about 1~GeV). However, no model to date has been able to reproduce the 
shape of $v_2(\eta)$. The results presented here for the centrality and energy 
dependence of this shape provide additional input that may give further 
insight into this issue.

\ack{
\small
This work was partially supported by U.S. DOE grants 
DE-AC02-98CH10886,
DE-FG02-93ER40802, 
DE-FC02-94ER40818,  
DE-FG02-94ER40865, 
DE-FG02-99ER41099, and
W-31-109-ENG-38, by U.S. 
NSF grants 9603486, 
0072204,            
and 0245011,        
by Polish KBN grant 2-P03B-062-27, and
by NSC of Taiwan under contract NSC 89-2112-M-008-024.
C.~M.~V. would like to thank the conference organizing committee for their 
financial support.}


\normalsize

\section*{References}
\begin{thebibliography}{15}

\bibitem{phflow130}
B.\ B.\ Back \emph{et al}, 
Phys.\ Rev.\ Lett. {\bf 89}, 222301 (2002).

\bibitem{Phobos_flow_200} 
B.\ B.\ Back \emph{et al.}, submitted to 
Phys.\ Rev.\ C Rapid Communications, (2004). nucl-ex/0704012

\bibitem{phobos_det}
B.\ B.\ Back \emph{et al},
Nucl.\ Instrum.\ Meth. A {\bf 499}, 603 (2003).

\bibitem{Poskanzer}
A.\ M.\ Poskanzer and S.\ A.\ Voloshin,
Phys.\ Rev.\ C {\bf 58}, 1671 (1998).

\bibitem{cmv} C.\ M.\ Vale,  Ph.\ D. thesis, Massachusetts Institute 
of Technology (2004).

\bibitem{Huovinen} P.\ F.\ Kolb, P.\ Huovinen, U.\ W.\ Heinz and H.\ 
  Heiselberg, Phys.\ Lett. {\bf500}, 232 (2001); 
  P.\ Huovinen., Private Communication (2004).

\bibitem{Star_v2_vsnpart_vspt_130} K.\ H.\ Ackermann \emph{et al.}, 
Phys.\ Rev.\ Lett.\ {\bf 86}, 402 (2001).

\bibitem{Phenix_v2_twopartcor_130} K.\ Adcox \emph{et al.}, 
Phys.\ Rev.\ Lett.\ {\bf 89}, 212301 (2002).

\bibitem{Star_v2_Pid_130} C.\ Adler \emph{et al.}, 
Phys.\ Rev.\ Lett.\ {\bf 87}, 182301 (2001).

\bibitem{Phenix_v2_pid_200} S.\ S.\ Adler \emph{et al.}, 
Phys.\ Rev.\ Lett.\ {\bf 91},182301  (2003).

\bibitem{BDO1} N.\ Borghini, P.\ M.\ Dinh, and J.-Y.\ Ollitrault, 
Phys.\ Rev.\, C {\bf 63}, 054906 (2001). 

\bibitem{BDO2} N. Borghini, P. M. Dinh, and J.-Y. Ollitrault. 
Phys.\ Rev.\, C {\bf 64}, 054901, (2001). 

\bibitem{Star_v2_Cumulant_130} C.\ Adler \emph{et al.}, 
Phys.\ Rev.\ C {\bf 66}, 034904 (2002).

\bibitem{KandT} Y.\ V.\ Kovchegov and K.\ L.\ Tuchin, 
Nucl.\ Phys.\, A {\bf 708}, 413 (2002).

\bibitem{phlimfrag}
B.\ B.\ Back, submitted to 
Phys.\ Rev.\ Lett., (2004). nucl-ex/0604021

\end{thebibliography}
\end{document}